\documentclass [18pt]{article}

\begin{document}
\textbf{Fundamental issues of quantum theory}

{Yeong-Shyeong Tsai}

{Department of Applied Mathematics, National Chung Hsing
University, Taichung, Taiwan}

\textbf{Abstract}

Though quantum theory is very successful in my respects, there are still
many serious problems such as infrared divergences and ultraviolet
divergences. It seems that perturbation theory and renormalization work
well. When the quantum field theory is introduced to generalize quantum
mechanics, there are two major changes or involutions, so to speak, of
quantum theory. One is second quantization and the other is that a system of
particles instead of single particle is considered. Most parts, if not all,
of quantum theory are based on the Lagrangian. In QFT, it is obvious that
second quantization relations imply that the constraints are introduced the
calculus of variation and Lagrange's multipliers must be adopted. In real
world a set or collection of any objects, such as a group of men or a pile
of stones, we can not expect they are equally likely. We can not assume that
they have the same weight. Therefore, there must be a distribution which is
associated with a system of particles. It means that a meaningful quantity
is adjoined in the system of particles. It seems that these concepts,
constraints and distribution are ignored in conventional approach. Further
more, there are two versions of quantization relations, one is prior to the
field equation and the other is posterior to the field equation. And it is
very difficult to find the posterior one. If the posterior one is found, of
course, these two versions of quantization must be the same one. Actually,
it implies that there is recursive problem. In this paper, we will discuss
these serious problems.

\textbf{Introduction}

We must notice that relations of the second quantization are constraints and
a system of particles instead of single particle is treated in QFT. From the
convention mathematical point of view, it must be considered in quantum
field theory that two major concepts should be included. One is the method
of the Lagrange's multipliers and the other is probability distribution.
Though it is very difficult job to construct a complete and perfect theory,
we have started to study such two topics. It was pointed out by Hung-Ming
Tsai et al in QTS3 conferences that the method of the Lagrange's multipliers
should be chosen to derive the field equation in gauge field theory. It was
pointed out by Hung-Ming Tsai et al in PASCOS conferences that regular basis
of field operators could resolve the soft infrared divergence of massless
mediators, such as Maxwell field, in QFT. The probability amplitude which is
introduced in quantum mechanics shall be discussed later. It was also
pointed out by Hung-Ming Tsai in SUSY 06 that the probability density
function of creators and annihilators in the Maxwell field operators are $1
/ E$ and it is not well-defined in $[0,\infty )$. Hence it will introduce
both infrared and ultraviolet divergences naturally. The fuzzy relation and
thermal factor are introduced in the same time to solve the ultraviolet
divergence. In this paper, we are going to study and summarize these two
topics in whole.

Consider a simple mathematical expression of quantum state vector

$\left| \phi \right\rangle = c_ + \left| { + z} \right\rangle + c_ - \left| {
- z} \right\rangle .$Here, $\left| {c_ + } \right|^2$and $\left| {c_ - }
\right|^2$ are interpreted as the probabilities that the electron in the
state $\left| \phi \right\rangle $ will be found in the states$\left| { + z}
\right\rangle $ and $\left| { - z} \right\rangle $ respectively. In order to
applying this interpretation to the field operator in QED, the field
operator $\Psi $ is rewritten as follows

\begin{equation}
\label{eq1}
\Psi = \sum\limits_\varepsilon {\sum\limits_{E = \varepsilon ,p}
{\sum\limits_s {(u_p^s b_p^s + \bar {v}_p^s d_p^{s\dag } )} } }
\end{equation}

Hence

\begin{equation}
\label{eq2}
\overline \Psi = \sum\limits_\varepsilon {\sum\limits_{E = \varepsilon ,p}
{\sum\limits_s {(u_p^s b_p^s + \bar {v}_p^s d_p^{s\dag } )} } } .
\end{equation}

Then the length of the vectors, $\bar {u}_p^s u_p^s $ and $\bar {v}_p^s
v_p^s $, must be associated with the distribution $1 / (1 + \exp ((E - \mu )
/ k_B T))$ for ferminons. Similarly, in the Maxwell field the associated the
distribution is $1 / ( - 1 + \exp ((E - \mu ) / k_B T))$. Since there are
many concerns such as Lorentz invariance and Maxwell's relations in
classical electrodynamics, it is a nontrivial work to introduce these
thermal factors, the distributions. In order to solve this problem, we adopt
some results of the fuzzy theory. A simple example will give a hint that
will lead to introduce the fuzzy theory into the quantum theory. The example
is: A class of girls is introduced to a class of boys. Probably, some of
them may get married. This probability behavior could be described by fuzzy
theory. Therefore, fuzzy relations are introduced between the class of field
operators and the vacuum state $\left| 0 \right\rangle $.

\textbf{The method Lagrange's multipliers}

Generally, most optimization must be solved by the method\textbf{
}Lagrange's multipliers. The simple elementary example of calculus will show
the result. Let $u = x^2 + 2x + 3y^4 - y + z^2 + z - 50$. In order to find
the relative extreme values of $u$, it will yield the same result when the
relations $xy = yx$, $yz = zy$ and $zx = xz$ are treated as constraints
though the constraints are ignored in usual computations. But in functional
calculus, we get new results when the imposed conditions or relations are
treated as constraints.

\textbf{The Hamiltonian operator and the Einstein's energy equation}

Based on the Lagrangian of a quantum system, the Hamilton's operator can be
obtained. The eigenvalue of Hamilton's operator is the energy of the system.
In gauge theory there are two kinds of fields. Each one is associated with a
class of particles. Therefore, there must be more energy operators. In
speaking of energy, the Einstein's energy equation, $E^2 = p^2 + m^2$, plays
an important role in beginning of developing the new field equation. Most
free field equations, Klein-Gordon equation and Dirac equation, are obtained
from the Einstein's energy equation. Since there is only energy which is
considered in quantum mechanics, the first quantization can give the energy
operator. And hence the energy of the system is the eigenvalue of this
energy operator. In gauge field theory there are at least two kinds of
particles. Each particle has its energy equation. Clearly, there are more
energy operators than before. Therefore, the field equations should be
modified in the interaction field. It is the Lagrange's multipliers that
make the field equations adjustable to fit the Einstein's energy equation.
The Einstein's energy equation must be satisfied no matter whether the
particles are free or not. In gauge field theory, the Dirac field equation
deviates from free one. In the gauge field theory, if we apply the method
Lagrange's multipliers, then we are able to obtain the same field equation
as free one. When energy is mentioned, it must be specified which energy,
the energy of the system, the energy of electron or the energy of photon, is
considered. The energy of the whole system must be derived from the
Lagrangian no matter whether the quantization relations are treated as
constraints or not. Now we quote some results of Hung-Ming Tsai in SUSY 06
to show how the method of Lagrange's multipliers applying in the gauge field
theory.

In order to study a simple gauge field theory, QED is taken as an example.
The Lagrangian is not stated here, the quantization relations are defined by
the differential equations.

Let

\begin{equation}
\label{eq3}
S_{ij}^M (x - x',t - t') = \{\Psi _i (x,t),\overline \Psi _j (x',t')\} \quad
.
\end{equation}

In order to solve the recursive problem, we must have some prior assumption.
From the result of free field, we have the prior information as such

\begin{equation}
\label{eq4}
(i\gamma ^\mu \partial _\mu - m)_{\alpha \gamma } S_{\gamma \beta }^M (x -
x',t - t') = 0,
\end{equation}

\noindent
with some initial condition

\begin{equation}
\label{eq5}
S_{ij}^M ({\rm {\bf x}} - {\rm {\bf x}}',0) = \delta _{ij} \delta ({\rm {\bf
x}} - {\rm {\bf x}}'),
\end{equation}

\noindent
and

\begin{equation}
\label{eq6}
\{\Psi _i (x,t),\Psi _j (x',t')\} = \{\overline \Psi _i (x,t),\overline \Psi
_j (x',t')\} = 0.
\end{equation}

By applying the method of Lagrange's multiplier, the field equation can be
obtained from the Lagrangian and the quantization relations

\begin{equation}
\label{eq7}
i\gamma ^\mu \partial _\mu \Psi - m\Psi + \Lambda \Psi = q\gamma ^\mu \Psi
A_\mu ,
\end{equation}

\noindent
where $\Lambda $ is a matrix of which the entries are the Lagrange's
multipliers. Now we are going to determine the entries of $\Lambda $.
Multiply both sides of the equation (\ref{eq7}) by $\overline \Psi (x',t')$ from the
left, multiply both sides of the equation (\ref{eq7}) by $\overline \Psi (x',t')$
from the right and take the sum of them. We get

\begin{equation}
\label{eq8}
(i\gamma ^\mu \partial _\mu - m + \Lambda )_{\alpha \gamma } S_{\gamma \beta
}^M = (q\gamma ^\mu A_\mu )_{\alpha \gamma } S_{\gamma \beta }^M .
\end{equation}

From (\ref{eq4}) and (\ref{eq8}), we get

\begin{equation}
\label{eq9}
\Lambda _{\alpha \gamma } S_{\gamma \beta }^M = (q\gamma ^\mu A_\mu
)_{\alpha \gamma } S_{\gamma \beta }^M .
\end{equation}

Setting $t$ to be $t'$ and taking the three dimensional integration of $x'$,
we get

\begin{equation}
\label{eq10}
\Lambda _{\alpha \beta } = (q\gamma ^\mu A_\mu )_{\alpha \beta } .
\end{equation}

\noindent
since $S_{\alpha \beta }^M ({\rm {\bf x}} - {\rm {\bf x}}',0) = \delta
_{\alpha \beta } \delta ({\rm {\bf x}} - {\rm {\bf x}}')$. Therefore, the
field equations are

\begin{equation}
\label{eq11}
(i\gamma ^\mu \partial _\mu - m)\Psi = 0,
\end{equation}

\noindent
and

\begin{equation}
\label{eq12}
(\frac{\partial ^2}{\partial x^2} + \frac{\partial ^2}{\partial y^2} +
\frac{\partial ^2}{\partial z^2} - \frac{\partial ^2}{\partial t^2})A^\mu =
q\overline \Psi \gamma ^\mu \Psi .
\end{equation}

Clearly, the energy of the electron which is obtained from the Dirac field
equation does not deviate from the Einstein's energy equation. We stress it
again that the energy operator of the whole system must be derived from the
Lagrangian, the Hamiltonian. When this operator apply to the vacuum state
$\left| 0 \right\rangle $, the energy of the energy of the system is
obtained. It is obvious the constraints, the quantization relation must be
obeyed during the computation process. It is of no doubt that the result is
the same as before, if the thermal factor is ignored. But the whole process
is more consistency in the sense of mathematics.

In order to describe the natural phenomenon, the distributions are
introduced in the relations between the field operators and the vacuum state
$\left| 0 \right\rangle $. When a non-constant polynomial of the whole field
operators $\Psi $ and $\overline \Psi $ operates on the vacuum state $\left|
0 \right\rangle $, there exists a fuzzy relation between the set of
operators, such as creators, annihilators and their products in this
polynomial, and the vacuum state $\left| 0 \right\rangle $. Hence this
operation yields a distribution which is inserted in the conventional
operation$\Psi \left| 0 \right\rangle $.

\textbf{Fuzzy Theory and the Distribution}

According to the spin of the particles, there are two
distributions which are associated with the particles. Let the
$T_f $ be defined\newline $T_f = \sqrt {1 / (1 + \exp ((E - \mu )
/ k_B T))} \sqrt {m / E} $. When the field $\Psi $ is spin $1 /
2$, from the fuzzy relation, the operation$\Psi \left| 0
\right\rangle $ shall yield the outcome result

\begin{equation}
\label{eq13}
\Psi \left| 0 \right\rangle = \sum\nolimits_p {\sum\nolimits_s
{\frac{N}{(2\pi )^{\raise0.7ex\hbox{$3$} \!\mathord{\left/ {\vphantom {3
2}}\right.\kern-\nulldelimiterspace}\!\lower0.7ex\hbox{$2$}}}T_f (b_p^s
\left| 0 \right\rangle u^s(p)\exp \varphi + d_p^{s\dag } \left| 0
\right\rangle v^s(p)\exp ( - \varphi )} } ).
\end{equation}

\noindent
where $\varphi = i({\rm {\bf p}} \cdot {\rm {\bf x}} - Et)$, $E = \sqrt {m^2
+ \left| {\rm {\bf p}} \right|^2} $and $N$ is a factor for normalization.

\textbf{The Parameter of System with Distribution}

When Hung-Ming Tsai, Po-Yu Tsai and Lu-Hsing Tsai was organizing the paper
for the Proceedings of the 3rd International Symposium on Quantum Theory and
Symmetries \textbf{QTS3}, Cincinnati, Ohio, 10-14 September 2003, they
discovered the distribution of a system of particles must be considered.
Therefore, the behavior of the system depends on some parameters such as
chemical potential and temperature. Therefore, most measured quantities must
be temperature dependent in QFT. In order to introduce these parameters, the
fuzzy relation is adopted. Due to the fuzzy relations between the field
operators and the vacuum state $\left| 0 \right\rangle $, the thermal
factors $\alpha \sqrt {1 / (1 + \exp ((E - \mu ) / k_B T))} $ is taken for
ferminons and $\beta \sqrt {1 / ( - 1 + \exp ((E - \mu ) / k_B T))} $ is
taken for bosons. Usually, $\alpha = 1$ for Dirac field, $\beta = 1 / \omega
_{\rm {\bf k}} $ for massless Maxwell field. The complete work are shown in
the 14th International Conference on Supersymmetry and the Unification of
Fundamental Interactions \textbf{SUSY `06}, University of California,
Irvine, California, June 12-17 2006. This is their major contributions in
quantum field theory since the ultraviolet divergence and infrared
divergence shall be removed by fuzzy relation. It seems that the work is
completed when quantization relations, the Lagrange's multipliers, fuzzy
relation and thermal factors are all put together. Actually, there some
problems must be studied since there is nothing to do with the uncertainty
principle in theory relativity and hence there is no uncertainty in the
energy equation. It will be all right in sense of uncertainty principle if
the life time of particles, electron, proton etc., are almost infinite, that
is, $\Delta t = \infty $. Hence that $\Delta E = 0$ is possible.

\textbf{Uncertainty Principle and the Mediators}

Based on the uncertainty principle, Yukawa predicted not only existence of
the pie mesons but also the mass of these mediators. Using the method of the
Lagrange's multipliers, we obtain the field equations in QED are

\begin{equation}
\label{eq14}
(i\gamma ^\mu \partial _\mu - m)\Psi = 0,
\end{equation}

\noindent
and

\begin{equation}
\label{eq15}
(\frac{\partial ^2}{\partial x^2} + \frac{\partial ^2}{\partial y^2} +
\frac{\partial ^2}{\partial z^2} - \frac{\partial ^2}{\partial t^2})A^\mu =
q\overline \Psi \gamma ^\mu \Psi .
\end{equation}

We find the equation (\ref{eq14}), the Dirac field equation, is different form of
the Einstein's energy equation, $E^2 = p^2 + m^2$, while the equation (\ref{eq15}),
the Maxwell equation, deviates from the Einstein's energy equation, $E^2 =
p^2 + m^2$. We argue that this term, $q\overline \Psi \gamma ^\mu \Psi $, in
the right hand side of equation (\ref{eq15}) is the balance term of the Einstein's
energy equation, $E^2 = p^2 + m^2$. Since life time of mediators is not
infinite, there must be some uncertainty in this equation.

\textbf{Conclusion and Discussion}

To construct a consistent theory is one thing, to verify this theory is
another thing. It takes a long time to verify the theory by experimental
results. Though the string theory is nice theory, there is no experimental
result to verify it so far. Hung-Ming Tsai, Po-Yu Tsai and Lu-Hsing Tsai
have predicted that most meaningful quantities derived in QFT such as energy
levels of hydrogen atoms, gyromegnetic ratio of electron and muon etc. might
be dependent on the temperature $T$ if the temperature is well defined. It
also takes long time to verify whether it is right or not.

In the Dirac field the thermal factor $\alpha \sqrt {1 / (1 + \exp
((E - \mu ) / k_B T))} $, $\alpha $ is a constant, $\alpha = 1$,
while in the Maxwell equation the thermal factor \newline $\beta
\sqrt {1 / ( - 1 + \exp ((E - \mu ) / k_B T))} $, $\beta $ is
momentum ${\rm {\bf k}}$ and hence energy $E$ dependent, $\beta =
1 / \omega _{\rm {\bf k}} $. If we force that $\beta $ is a
constant, $\beta = 1$, then the fractional operator might be
introduced. Let the operator $D_t $ be defined $D_t = \partial /
\partial t$. Let $D^{1 / 2}D^{1 / 2} = D_t $. Obviously, we are
going to adopt the fractional functional calculus since $D^{1 /
2}$ will be introduced into QFT.

Let the Maxwell field be $A^\mu $. Let $A_{1 / 2}^\mu = D^{1 /
2}A^\mu $. Ignoring the gauge fixed, radiation gauge or Lorentz
gauge, $A_{1 / 2}^\mu $ satisfy the Maxwell field equation, the
wave equation. Roughly speaking, in gauge field theory, $A^\mu $
can be replaced by$A_{1 / 2}^\mu $ since the invariance of local
gauge transformation implies there must be some adjoin field
$A^\mu $ which, the symbols in sense of mathematical operation,
collaborates with gauge transformation. The new field satisfies
the wave equation. Clearly, the adjoin field (symbol) is not
unique. The new field $A_{1 / 2}^\mu $ is chosen instead of $A^\mu
$, then there will be no infrared divergence if $\beta $ in $\beta
\sqrt {1 / ( - 1 + \exp ((E - \mu ) / k_B T))} $ is a constant,
$\beta = 1$. Since some problems, such as the covariance in the
Lorentz transformation and homogenous dimension of each term in
the Lagrangian and the field equation must be solved, some new
matrix must be introduced in the Lagrangian. This is not a simple
work. Fractional calculus is very popular in the applied
mathematics. It takes time to develop the new theory of QFT by
fractional functional calculus.

In this paper, we try to make quantum field theory more consistent in the
senses of mathematics and physics. It is not an occasional
 case to introduce the thermal factor in constrained QFT because
most distributions of particles are derived by optimizing an
 objective function with constraints. These constraints are: the
total energy of the system of particles is equal to a finite
constant and the total number of the system of particles is equal
to a finite constant. Finally, the temperature and chemical
potential of the system are obtained directly or indirectly from
the Lagrange's multipliers of these constraints.

\textbf{References}

1. Yeong-Shyeong Tsai, Hung-Ming Tsai, Po-Yu Tsai and Lu-Hsing Tsai, the
Proceedings of the 3rd International Symposium on Quantum Theory and
Symmetries (\textbf{QTS3}), Cincinnati, Ohio, 10-14 September 2003, World
Scientific, Singapore (2004), pp. 431-436.

2. Yeong-Shyeong Tsai, Hung-Ming Tsai, Po-Yu Tsai and Lu-Hsing Tsai, the
Proceedings of the 10th International Symposium on Particles, Strings and
Cosmology (\textbf{PASCOS `04}), Northeastern University, Boston,
Massachusetts, August 16-22, 2004, World Scientific, Singapore (2005), pp.
549-553.

3. Yeong-Shyeong Tsai, Lu-Hsing Tsai, Hung-Ming Tsai and Po-Yu Tsai,
"\textit{Recursive Relation Between the Field Equations and the Quantization of the Field}," 14th International Conference on Supersymmetry and the Unification of
Fundamental Interactions(\textbf{SUSY `06}), University of California,
Irvine, California, June 12-17 2006.

4. A. Yariv\textit{, An Introduction to Theory and Application of Quantum Mechanics, }John Wiley {\&} Son$, $(1982).

5. H. -J. Zimmermann, \textit{Fuzzy set theory and its applications}, 2nd ed., Kluwer Academic Publishers, Boston, (1991).

6. D. Lurie, \textit{Particles and fields}, Wiley, 1968.

7. T. Muta, \textit{Foundations of Quantum Chromodynamics}, 2nd ed., World Scientific, Singapore (1998)

\end{document}